\documentclass{svjour3}                     
\smartqed  

\usepackage{graphicx}
\usepackage[sort, compress]{cite}

\journalname{Computational Mechanics}

\begin{document}

\title{Compressible air flow through a collapsing liquid cavity}

\author{Stephan Gekle \and Jos\'e Manuel Gordillo}

\institute{S.~Gekle \at
              Department of Applied Physics and J.M. Burgers Center for Fluid Dynamics, University of Twente, P.O. Box 217, 7500 AE Enschede, The Netherlands\\
              \email{s.gekle@tnw.utwente.nl}           
           \and
           J.~M.~Gordillo \at
              Area de Mec\'anica de Fluidos, Departamento de Ingener\'ia Aeroespacial y Mec\'anica de Fluidos, Universidad de Sevilla, Avenida de los Descubrimientos s/n, 41092 Sevilla, Spain\\
              \email{jgordill@us.es}
}

\date{Received: date / Accepted: date}

\maketitle

\begin{abstract}
We present a multiscale approach to simulate the impact of a solid object on a liquid surface: upon impact a thin liquid sheet is thrown upwards all around the rim of the impactor while in its wake a large surface cavity forms. Under the influence of hydrostatic pressure the cavity immediately starts to collapse and eventually closes in a single point from which a thin, needle-like jet is ejected. Existing numerical treatments of liquid impact either consider the surrounding air as an incompressible fluid or neglect air effects altogether. In contrast, our approach couples a boundary-integral method for the liquid with a Roe scheme for the gas domain and is thus able to handle the fully \emph{compressible} gas stream that is pushed out of the collapsing impact cavity. Taking into account air compressibility is crucial, since, as we show in this work, the impact crater collapses so violently that the air flow through the cavity neck attains supersonic velocities already at cavity diameters larger than 1~mm. Our computational results are validated through corresponding experimental data.
\end{abstract}

\section{Introduction}


The thin jet ejected after the impact of an object on a liquid surface has been one of the icons of fluid mechanics since the days of Worthington more than a century ago \cite{Worthington_book_1908}. Since then a fair amount of computational studies on the impact of liquid drops \cite{OguzProsperetti_JFM_1990, MortonRudmanLiow_PhysFluids_2000, JosserandZaleski_PhysFluids_2003, LeneweitEtAl_JFM_2005, ZhengEtAl_JComputPhys_2005, CoyajeeBoersma_JComputPhys_2009} or solid objects \cite{GreenhowMoyo_PhilTransRoySocLondonA_1997, Gaudet_PhysFluids_1998, BattistinIafrati_JFluidStruct_2003, LiEtAl_JComputPhys_2005, VellaMetcalfe_PhysFluids_2007, Lin_ComputFluids_2007, GekleEtAl_PRL_2008, GekleEtAl_PRL_2009, BergmannEtAl_JFM_2009, DoQuangAmberg_PhysFluids_2009, GekleEtAl_PRL_2010} has been reported. In these works a number of different methods have been employed including Arbitrary-Langrangian-Eulerian, Boundary-Integral, Volume-of-Fluid, Level-Set, or combinations thereof. Despite this variety a common feature is that the dynamics of the surrounding air was either neglected altogether \cite{OguzProsperetti_JFM_1990, GreenhowMoyo_PhilTransRoySocLondonA_1997, Gaudet_PhysFluids_1998, MortonRudmanLiow_PhysFluids_2000, BattistinIafrati_JFluidStruct_2003, LiEtAl_JComputPhys_2005, VellaMetcalfe_PhysFluids_2007, Lin_ComputFluids_2007, GekleEtAl_PRL_2008, GekleEtAl_PRL_2009, BergmannEtAl_JFM_2009} or it was treated as an incompressible fluid \cite{JosserandZaleski_PhysFluids_2003, LeneweitEtAl_JFM_2005, ZhengEtAl_JComputPhys_2005, DoQuangAmberg_PhysFluids_2009, CoyajeeBoersma_JComputPhys_2009} in the regime for low Mach numbers.

\begin{figure}
\includegraphics[width=0.24\columnwidth]{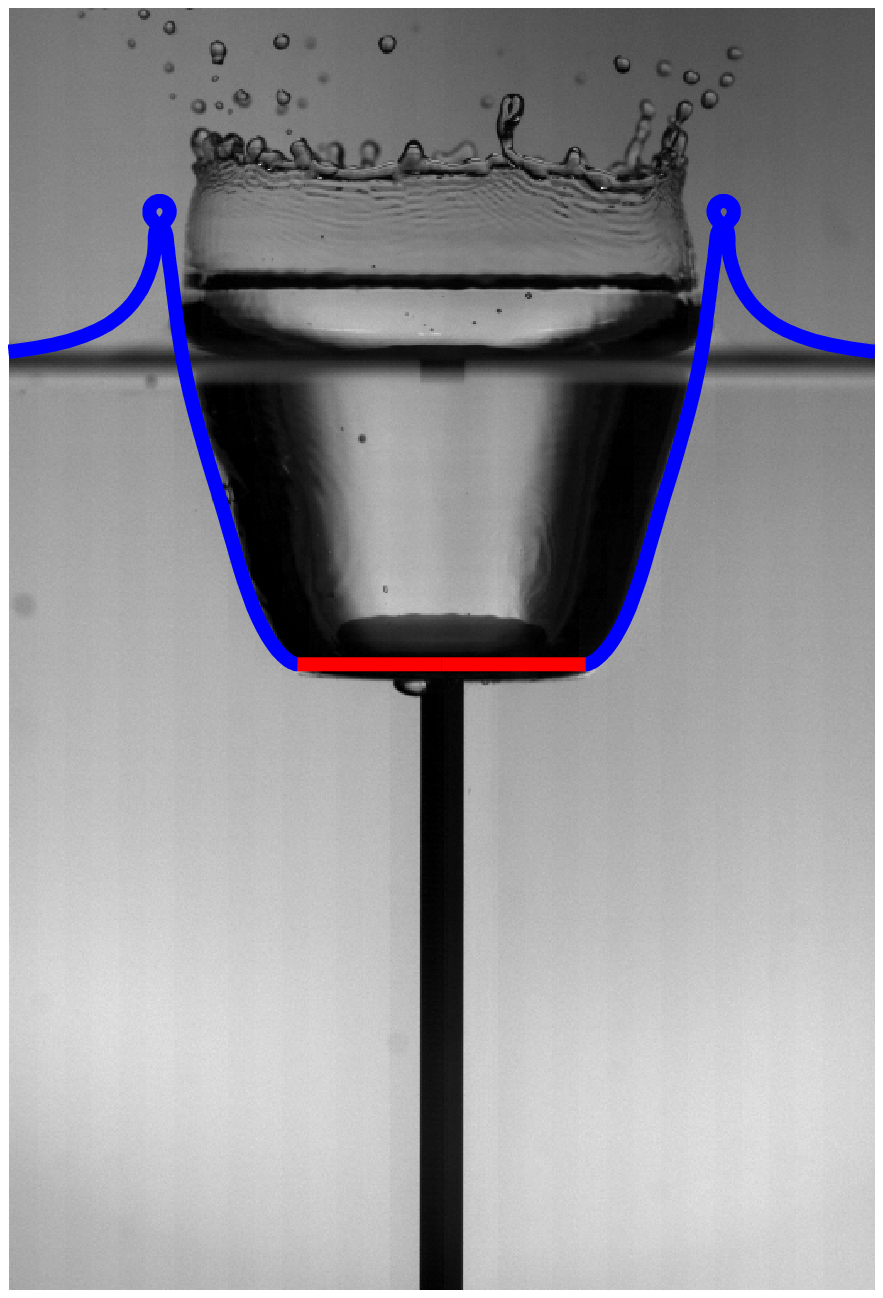}
\includegraphics[width=0.24\columnwidth]{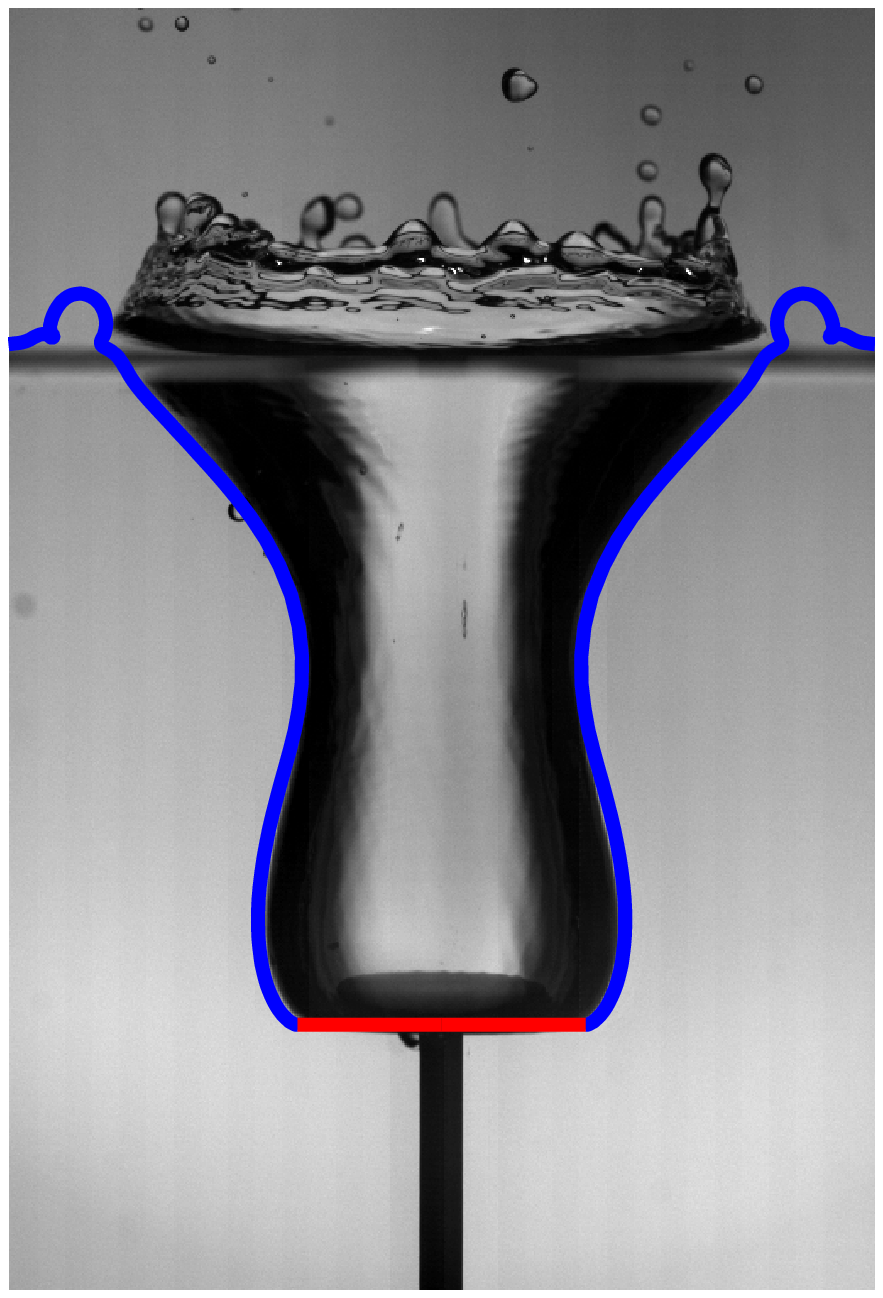}
\includegraphics[width=0.24\columnwidth]{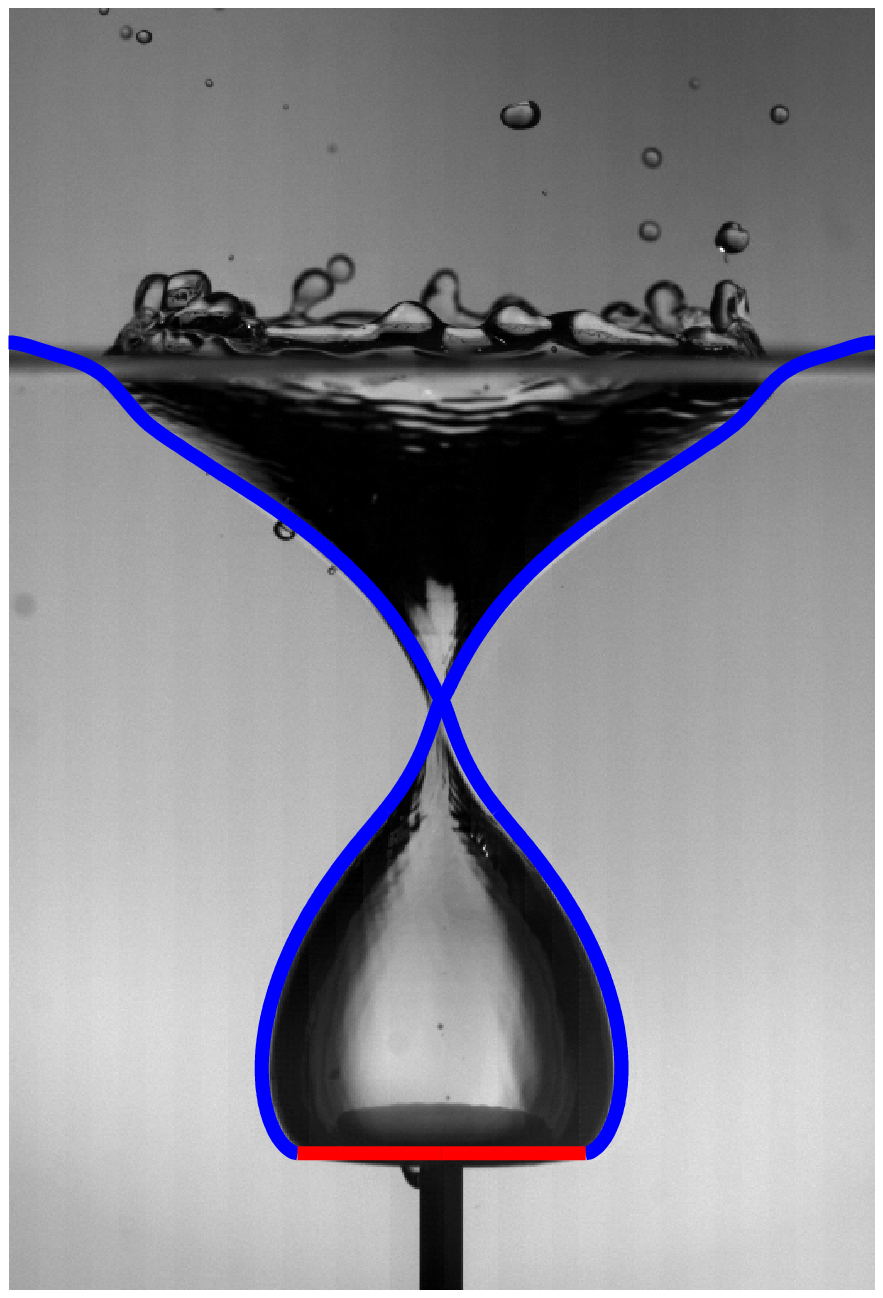}
\includegraphics[width=0.24\columnwidth]{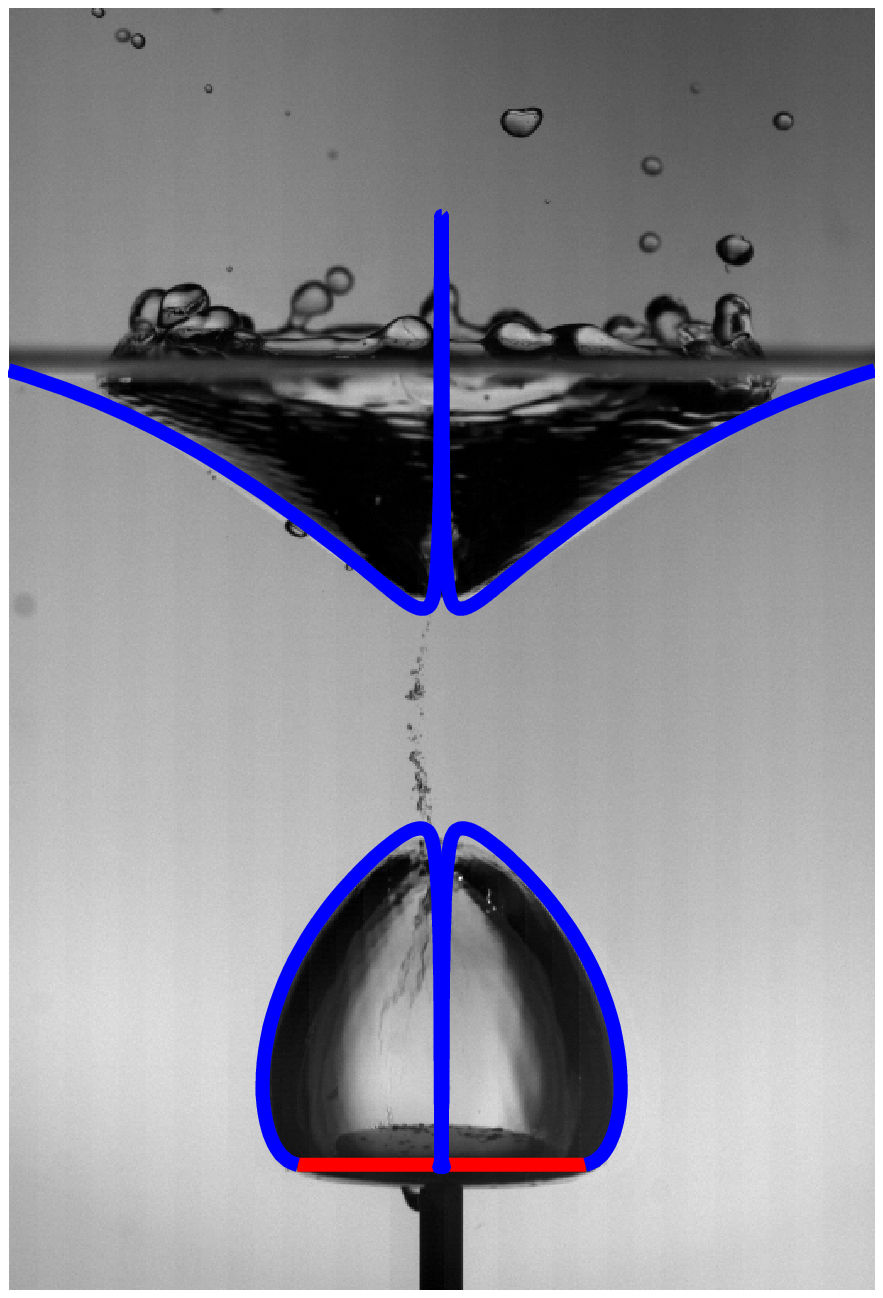}
\caption{The sequence of events as a circular disc of 2~cm radius impacts a water surface at 1~m/s: (a) Immediately after impact a crown splash is thrown up into the air and an impact cavity forms below the free surface. The moving disc draws air into the cavity. (b) Hydrostatic pressure pushes the cavity together and the direction of air flow reverses. (c) Eventually the cavity closes in a single point. (d) After closure two violent jets are ejected up- and downwards from the closure location. The blue and red lines represent the free surface and the disc, respectively, from the present numerical simulation with air in (a)-(c). For the jetting in (d) the single-phase simulation from \cite{GekleEtAl_PRL_2009} is shown.}
\label{fig:surfaceProfiles}
\end{figure}

In this work we report a seemingly simple and harmless situation which nevertheless requires to model the dynamics of the gas phase in a fully compressible way: a circular disc of 2~cm radius which impacts on a liquid surface with a speed of 1~m/s. Figure~\ref{fig:surfaceProfiles} illustrates the sequence of events during the disc impact extracted from high-speed video images \cite{BergmannEtAl_JFM_2009, GekleEtAl_PRL_2010} and compared to the results of our simulations. Upon impact, first a thin liquid splash is thrown up all around the circumference of the penetrating disc. In the wake of the impactor a large cavity is created which subsequently starts to collapse due to the hydrostatic pressure of the surrounding liquid. When the cavity closes about half-way down its length two very fast and thin jets are observed shooting up- and down from the closure point \cite{BartoloJosserandBonn_PRL_2006, DengAnilkumarWang_JFM_2007, GekleEtAl_PRL_2009, GekleGordillo_preprint}.

In the beginning of the process, obviously, air is drawn into the cavity by the moving disc. At a later stage, however, this inward flow is counteracted by the shrinking of the cavity volume itself and the direction of air flow is not a priori clear. We will show that in the competition between cavity expansion (just above the disc) and shrinking (around the neck), eventually the shrinking becomes dominant. Accordingly, the air flow through the neck reverses and air is pushed out of the cavity. The collapse is so violent that the air stream can attain \emph{supersonic} speeds. To handle this situation, we implement an axisymmetric boundary-integral method (BIM) to simulate the motion of the liquid surface which is two-way coupled with a fully compressible Roe solver for the highly unsteady gas flow.

In Section \ref{sec:methods} we will introduce briefly our BIM and Roe implementations and then describe in some detail the two-way coupling between the BIM for the liquid and the Roe scheme for the gas domain. Sections \ref{sec:justification} and \ref{sec:structure} show the main physical results extracted from our simulations and Section~\ref{sec:validation} briefly summarizes the experimental validation carried out in \cite{GekleEtAl_PRL_2010}. Section~\ref{sec:conclusions} concludes the article.

%
%
%
%

\section{Numerical methods}
\label{sec:methods}

In the impact process viscous effects are negligible as can be seen by estimating the Reynolds numbers for gas and liquid as $\mathrm{Re}_{g,l} = V_0R_0/ \nu_{g,l}$. Here, $V_0=1$~m/s is the impact speed, $R_0=2$~cm is the disc radius, and $\nu_g=1.46\cdot 10 ^{-5}\;\mathrm{m^2/s}$ and $\nu_l = 1.12\cdot 10 ^{-6}\;\mathrm{m^2/s}$ are the dynamic viscosities of gas and liquid at 15$^\circ$C, respectively. Both $\mathrm{Re}_g$ and $\mathrm{Re}_l$ are larger than $10^3$ and viscosity is thus negligible. Furthermore, we showed in earlier works \cite{BergmannEtAl_JFM_2009} that only a negligible amount of vorticity is present in the system. We can thus assume the flow to be inviscid and irrotational as is required for the boundary-integral method and the applicability of the inviscid Euler equations.

Our simulation is split in two stages. In the first, incompressible stage, a two-fluid boundary integral method is used to simulate the first part of the impact and cavity collapse where gas velocities are of the order of the disc velocity. In the second, compressible stage, we use a single-fluid BIM for the liquid coupled to a Roe scheme to solve the compressible Euler equations in the gas domain \cite{ChengLiu_JHydrodynamicsB_2007}.


The transition between both stages is fixed at the moment that the air flow through the cavity neck reverses and the gas axial velocity at the neck is zero, i.e. $u_\mathrm{neck}=0$.
\footnote{The transition point can of course also be taken somewhat later as long as gas velocities are still low enough to neglect compressibility. We tried $u_\mathrm{neck} =$ 10, 20, and 50~m/s which all give similar results.}

At the transition moment the liquid cavity has an elongated shape with a neck located roughly at the middle as can be seen in Fig.~\ref{fig:surfaceProfiles}~(b). Furthermore, the gas flow is to a very good approximation one-dimensional directed along the symmetry axis of the cavity (as can be verified by flow profiles obtained from the two-fluid BIM shown in Section~\ref{sec:justification}). This makes the situation reminiscent of gas flow through a converging-diverging de-Laval nozzle frequently encountered in aerodynamics \cite{Laney_book_1998}. In this spirit, we remove the inner potential fluid  during the compressible stage and replace it with a compressible gas described by the one-dimensional Euler equations. To integrate Euler's equations in time we use the well-known scheme due to Roe \cite{Roe_JComputPhys_1981}. This description is valid along most of the cavity. Above the initial free surface as well as in a small zone above the disc the air dynamics can be neglected as will be described in Section~\ref{sec:coupling}. The pronounced difference to the standard nozzle situation, however, is that in our case the ``nozzle'' geometry is determined by a \emph{liquid} interface which is changing rapidly in time and thus creates a highly unsteady gas flow.

The two-way coupling between the gas and the liquid domains is accomplished via (i) the interfacial shape and its instantaneous velocity which is provided by the BIM and serves as an input into the gas solver and (ii) the gas pressure which is obtained from the solution of the Euler equations and serves as a boundary condition for the BIM. Above the location of the initial water level the surface pressure of the BIM remains atmospheric. Our combined BIM/Euler method has the advantage that we retain the computational efficiency of the BIM which requires only the modeling of the boundary and not of the entire liquid domain in contrast to, e.g.~compressible air/water treatments based on the Volume-of-Fluid method for sloshing tanks \cite{GodderidgeEtAl_OceanEng_2009, ChenPrice_PhysFluids_2009}.


All quantities are non-dimensionalized with the disc radius $R_0=2$~cm and the impact velocity $V_0=1$~m/s. As a third quantity for non-dimensionalization we use the density of water $\rho_l=998.23$~kg/m$^3$ for the equations concerning the liquid domain and the density of air (at rest under atmospheric pressure) $\rho_g=1.2$~kg/m$^3$ for the gas equations. Since viscosity is neglected we have four dimensionless parameters which are the Froude number, the Weber number, the Euler number for the liquid, and the Euler number for the gas defined as:
\begin{eqnarray}
\mathrm{Fr} = \frac{V_0^2}{gR_0}\\
\mathrm{We} = \frac{\rho_l R_0V_0^2}{\sigma}\\
\mathrm{Eu}_l = \frac{p_a}{\rho_l V_0^2}\\
\mathrm{Eu}_g = \frac{p_a}{\rho_g V_0^2}.
\end{eqnarray}
with the atmospheric pressure $p_a=101.3$~kPa, the acceleration of gravity $g=9.81$~m/s$^2$, and the air/water surface tension $\sigma=72.8$~mN/m. We use cylindrical coordinates $r$ and $z$ with $z=0$ at the height of the initial free surface as is appropriate for our axisymmetric setup. In all computations the disc is assumed to have zero thickness.

The boundary-integral method for a single fluid (liquid) is briefly sketched in \ref{sec:BISingle} and its extension to two fluids (liquid and air) is given in \ref{sec:BITwo}. Some important aspects of our specific BIM implementation are described in \ref{sec:BISpecific}. The Roe solver is, again briefly, presented in Section~\ref{sec:Roe}. In Section~\ref{sec:coupling} we describe the coupling between the gas and liquid domains during the compressible stage.



\subsection{Boundary integral formulation}

\subsubsection{Boundary integral method for a single fluid}\label{sec:BISingle}

If liquid flow is inviscid and irrotational as is the case in our setup \cite{BergmannEtAl_JFM_2009, GekleEtAl_PRL_2009} the flow field $\vec{v}$ can be described as the gradient of a scalar potential $\phi$
\begin{equation}
\vec{v}=\nabla\phi
\end{equation}
which satisfies Laplace's equation throughout the liquid domain
\begin{equation}
\Delta \phi = 0. \label{eqn:Laplace}
\end{equation}
From Eq.~(\ref{eqn:Laplace}) the boundary-integral equation can be derived using Green's identities \cite{Pozrikidis_book_1997}:
\begin{equation}
\beta\phi\left(\vec{r}\right) = \int_S \left[\frac{1}{\left|\vec{r}-\vec{r}\,'\right|}\phi_n\left(\vec{r}\,'\right) - \phi\left(\vec{r}\,'\right)\vec{n}\cdot\nabla'\frac{1}{\left|\vec{r}-\vec{r}\,'\right|}\right]dS'
\label{eqn:BI}
\end{equation}
with $S$ denoting the boundary, $\vec{n}$ the normal vector pointing out of the liquid domain, and $\phi_n=\vec{n}\cdot\nabla\phi$ the normal derivative of the potential along the boundary. Equation~(\ref{eqn:BI}) expresses the potential $\phi$ at an arbitrary point $\vec{r}$ inside the domain (then $\beta=4\pi$) or on its edge (then $\beta=2\pi$) merely in terms of quantities that are defined on the surface of the liquid domain. This has the major advantage that it is sufficient to evolve the surface quantities $\phi$ and $\phi_n$ effectively reducing the computational problem by one spatial dimension.

If on every point along the interface either $\phi$ or $\phi_n$ is known, Eq.~(\ref{eqn:BI}) can be solved for the missing quantity such that afterwards both $\phi$ and $\phi_n$ are available along the entire boundary. With this the interfacial velocity can be computed as
\begin{equation}
\vec{v} = \frac{\partial\phi}{\partial s}\vec{t} + \phi_n\vec{n}
\end{equation}
with $\partial/\partial s$ denoting the tangential derivative and $\vec{t}$ the tangential vector along the surface. The numerical procedure for this is sufficiently well described in the literature \cite{BlakeTaibDoherty_JFM_1986, OguzProsperetti_JFM_1993, Pozrikidis_book_1997} and the details of the present implementation will be given in Section~\ref{sec:BISpecific}.

The boundary conditions for solving Eq.~(\ref{eqn:BI}) are provided as follows: on the disc the liquid is required to follow the disc's motion meaning that $\phi_n=-1$ in non-dimensional coordinates. On the air/liquid interface the potential $\phi$ is specified by integrating Bernoulli's equation:
\begin{equation}
\frac{D\phi}{Dt} = \frac{1}{2}\left|\vec{v}\right|^2 - \mathrm{Eu}_l\left(p-1\right) - \frac{1}{\mathrm{Fr}}z -\frac{1}{\mathrm{We}}C \label{eqn:Bernoulli}
\end{equation}
with $D/Dt=\partial/\partial t +\vec{v}\cdot\nabla$ denoting the material derivative, $C=\nabla\cdot\vec{n}$ the local curvature of the interface, and $p=p_g/p_a$ the dimensionless gas pressure. Here we assume that the pressure at a point far away from the symmetry axis where the flow is quiescent is atmospheric, i.e.~1 in non-dimensional coordinates.

In total, the boundary-integral simulation for a single fluid contains three substeps to advance from time step $j$ to $j+1$ \cite{BlakeTaibDoherty_JFM_1986, OguzProsperetti_JFM_1993, Pozrikidis_book_1997}. First, with the velocity $v^{(j)}$ we integrate Bernoulli's equation~(\ref{eqn:Bernoulli}) to compute the potential $\phi^{(j+1)}_f$ along the free surface. The second step updates the position of the free surface by integrating
\begin{equation}
\frac{d\vec{r}}{dt} = \vec{v}\label{eqn:kinematic}
\end{equation}
and moves the disc downwards with its prescribed velocity. Finally, we solve the boundary-integral equation (\ref{eqn:BI}) to obtain the liquid potential over the disc $\phi^{(j+1)}_d$ and the normal derivative of the potential over the free surface $\phi_{n,f}^{(j+1)}$.  Then the process repeats.

\subsubsection{Boundary-integral method for two fluids}\label{sec:BITwo}

The BIM can be extended to describe two immiscible fluids with a moving interface separating both phases. Our approach closely follows that of \cite{RodriguezRodriguezGordilloMartinezBazan_JFM_2006, GordilloSevillaMartinezBazan_PhysFluids_2007} and is thus only briefly sketched here. The liquid and gas phase both satisfy the boundary-integral equation (\ref{eqn:BI}) within their respective domains. At the interface between the two fluids tangential stresses vanish since the fluids are inviscid, i.e., the fluids can slip freely along the boundary. To ensure continuity of the interface, however, the normal velocities of both phases must exactly balance
\begin{equation}
\phi_n = - \phi_{n,g}
\label{eqn:contPhin}
\end{equation}
with $\phi_{n,g}$ being the normal derivative of the gas potential and $\phi_n$ that of the liquid as defined above. The minus sign is due to the normal vector pointing always out of the respective domains. Furthermore, the pressure jump across the interface is given by the Laplace pressure.
These two conditions are sufficient to derive the two-fluid version of the BIM \cite{RodriguezRodriguezGordilloMartinezBazan_JFM_2006, GordilloSevillaMartinezBazan_PhysFluids_2007}.

\subsubsection{Specific implementation of the boundary-integral method}\label{sec:BISpecific}

In our axisymmetric situation the surface integrals in Eq.~(\ref{eqn:BI}) can be reduced to one-dimensional line integrals in the $(r,\,z)$-plane after analytical integration over the azimuthal angle. Numerical integration is carried out using 8-point Gaussian quadrature with the weak logarithmic singularities removed analytically as in \cite{OguzProsperetti_JFM_1993}. Between the nodes the interface shape and potentials are interpolated using cubic splines with the node-to-node distance serving as the spline parameter $s$. Natural boundary conditions (i.e.~a vanishing second derivative with respect to $s$) are used for all splines, except for the potential at the connection between the disc's edge and the free surface as described below. Once the known integrals in Eq.~(\ref{eqn:BI}) are evaluated we transform Eq.~(\ref{eqn:BI}) into a matrix equation which is solved by LU decomposition \cite{BlakeTaibDoherty_JFM_1986,OguzProsperetti_JFM_1993}.

Time-stepping for the integration of Eqs.~(\ref{eqn:Bernoulli}) and (\ref{eqn:kinematic}) is carried out by an iterative Crank-Nicholson procedure during the incompressible stage. In the compressible stage, we use a simpler forward-Euler scheme for ease of coupling between the BIM and the Roe solver in the two respective domains. The time step in the incompressible stage is determined by the condition that neighboring nodes may not collide even if their velocities were directed exactly towards each other. This leads to:
\begin{equation}
\Delta t' = f\cdot\mathrm{min}_i\left(d_i/v_i\right)
\end{equation}
with $i$ running over all nodes, $v_i$ the free surface velocity at node $i$, and $d_i$ the distance to the neighboring node. This quantity is multiplied with a safety factor $f$ which is in most simulations chosen to be $0.1$. In the compressible stage the time-step is determined by the stability condition of the Roe solver as described in the next section.

To ensure a continuous boundary of the liquid domain the last node of the free surface remains fixed at the disc's edge. This connection point requires some special consideration. First, the surface is not smooth and thus the prefactor $\beta$ in Eq.~(\ref{eqn:BI}) must be modified to become $\beta=2\alpha$ for the liquid and $\beta=2\pi - 2\alpha$ for the gas equations. Here, $\alpha$ is the angle connecting the horizontal disc and the tangent to the free surface at the connection point through the liquid domain (i.e.~$\alpha>\pi$ in our situation). Second, the connection point is at the same time the last node, $\mathcal{F}$, of the free interface and the first node, $\mathcal{D}$, of the disc. To solve Eq.~(\ref{eqn:BI}) we consider it part of the disc, i.e.~we impose $\phi_n=-1$ and obtain the corresponding value for $\phi$. Together with the pinning at the disc's edge, the position $\vec{r}$ and the liquid velocity $\vec{v}$ at $\mathcal{D}$ are thus completely determined. We then need to ensure that $\mathcal{F}$ and $\mathcal{D}$ remain identical. For this we first copy the spatial coordinates of $\mathcal{D}$ on $\mathcal{F}$. To ensure further that also the velocity $\vec{v}$ is identical on both nodes we project $\vec{v}$ on the tangential and normal vectors of the free surface at $\mathcal{F}$ which determines the values of $\phi_n$ and $\partial\phi / \partial s$. The latter is imposed as a boundary condition on the spline function for $\phi$. This procedure ensures that the velocity $\vec{v}$ of the connection point is identical when seen from the disc or from the free surface, even though the respective normal and tangent vectors are discontinuous.

In our simulations the free surface extends from the edge of the disc at $r=1$ out to 100 where any motion is negligible and the surface is cut off. Note that BIMs do not require a closed liquid domain since the portion of $S$ at infinity gives no contribution to Eq.~(\ref{eqn:BI}) provided that $\phi$ goes to zero there. We use an adaptive mesh to ensure that the sensitive areas such as the crown splash or the cavity neck are properly resolved without wasting computation time by placing a large amount of nodes on unimportant parts. During the incompressible stage the local node distance $d$ is inversely proportional to the local curvature $C$ with a proportionality constant between 0.05 and 0.005. We impose a maximum distance ($d_\mathrm{max}=10$) and minimum distance ($d_\mathrm{min}=0.01$) and construct our regridding algorithm such that large gradients in the node density, which might cause instabilities, are avoided. When coupling with the Euler solver during the compressible stage, the BI mesh corresponds to the grid cells used for the Roe scheme as will be described in Section~\ref{sec:coupling}. Note that then we also allow for node distances smaller than $d_\mathrm{min}$.

\begin{figure}
\begin{center}
\includegraphics[width=0.5\columnwidth]{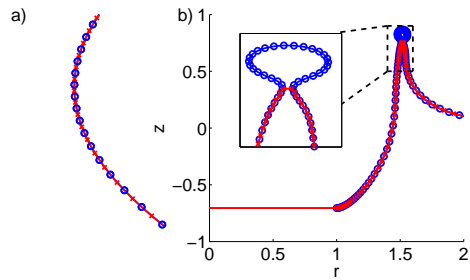}
\caption{(a) Illustration of the regridding procedure: every $n$ time steps the surface is reconstructed with the new nodes (red crosses) shifted to lie halfway between the old nodes (blue circles) in order to avoid amplification of small numerical disturbances. (b) The thin liquid sheet ejected after impact breaks up into droplets which are cut off and discarded as illustrated by the blue line (circles) before and the red line (crosses) after cut-off.}\label{fig:grid}
\end{center}
\end{figure}

Boundary-integral methods are known to be vulnerable to instabilities (see e.g.~\cite{OguzProsperetti_JFM_1990}) due to the lack of a naturally damping viscosity which prevents small numerical disturbances from building-up over time. To handle such instabilities we use a smoothing algorithm as follows: At every $n$th (usually $2\le n\le 10$) time step the free surface nodes are redistributed such that new nodes fall exactly half-way between old nodes as illustrated in Fig.~\ref{fig:grid}~(a). This periodic regridding procedure efficiently ensures the stability of the numerical scheme \cite{OguzProsperetti_JFM_1990}.

A specific detail of our physical problem is the thin sheet of liquid thrown up around the rim of the impacting disc (see Fig.~\ref{fig:surfaceProfiles}~(a)). In reality, this sheet will quickly develop non-axisymmetric instabilities leading to the formation of individual droplets earning it the title of a ``crown splash'' as seen, e.g., in the famous pictures of \cite{Worthington_book_1908}. As here we are not interested in the details of this splash and our axisymmetric code is not able to handle the droplet formation in any case, we cut off the splash as soon the distance between the two sides of the liquid sheet at a given point falls below the local node distance as illustrated in Fig.~\ref{fig:grid}~(b). The actual surface surgery is similar to the one used in \cite{GekleEtAl_PRL_2009} to handle the pinch-off of the cavity prior to jet formation.

\subsection{Roe method for a compressible gas}\label{sec:Roe}

In the compressible stage the inner gas is described by the one-dimensional Euler equations for conservation of mass, momentum, and energy:
\begin{eqnarray}
\frac{\partial \left(\rho S\right)}{\partial t} + \frac{\partial\left(\rho u S\right)}{\partial z} & = & 0\label{eqn:mass}\\
\frac{\partial \left(\rho u S\right)}{\partial t} + \mathrm{Eu}_g\frac{\partial \left(pS\right)}{\partial z} + \frac{\partial\left(\rho u^2S\right)}{\partial z} & = & \mathrm{Eu}_gp\frac{\partial S}{\partial z}\label{eqn:momentum}\\
\frac{\partial\left(\rho E S\right)}{\partial t} + \frac{\partial \left(\rho uHS\right)}{\partial z} &=& -\mathrm{Eu}_g p\frac{\partial S}{\partial t}.
\label{eqn:energy}
\end{eqnarray}
All quantities are dimensionless and $S$ is the cross-sectional area of the cavity, $\rho$ the gas density, $u$ the velocity, and $p$ the gas pressure. The total energy $E$ per unit mass is defined as
\begin{equation}
E = \frac{1}{\gamma-1}\frac{p}{\rho} + \frac{1}{2}u^2
\end{equation}
and the total enthalpy $H$, again per unit mass, is
\begin{equation}
H = E + \frac{p}{\rho}.
\end{equation}
The fact that we use $p_a$ as a reference pressure for $p$ leads to the appearance of $\mathrm{Eu}_g$ in Eqs.~(\ref{eqn:mass})--(\ref{eqn:energy}). Note the two source terms $p\partial S / \partial z$ and $-p\partial S / \partial t$ on the right-hand side which account for a cavity radius which is changing in space and time.

Integration of Eqs.~(\ref{eqn:mass})--(\ref{eqn:energy}) is carried out using a Roe scheme \cite{Roe_JComputPhys_1981, Laney_book_1998} whose implementation is fairly standard and thus omitted here.

The time step during the compressible stage is restricted by the Courant-Friedrich-Lewy (CFL) stability condition for the Roe solver. We fix a constant CFL number $C^*<1$ (usually $C^*=0.5$) and determine the time step by:
\begin{equation}
\Delta t = C^* \frac{\Delta z}{\mathrm{max}_i\left(\left|u_i\right|+c_i\right)}
\end{equation}
with the cell size $\Delta z$, the local speed of sound in each cell $c_i$, and the index $i$ running over all cells.

One detail that is of interest are the boundary conditions at the upper and lower end of the cavity. In contrast to the standard problem of flow through a nozzle which possesses an inlet on one side and an outlet on the other side, in our case gas leaves the shrinking cavity on both sides. Since both outflows are subsonic we can prescribe one flow quantity at each boundary \cite{Laney_book_1998}. To compute the flux through the lower (upper) face of the first (last) computational cell we add a boundary cell at each end whose state values are calculated as follows.

At the upper exit we impose that the pressure in the boundary cell be atmospheric. In order to compute the density and velocity of the boundary cell, let $\rho_N$, $u_N$, and $p_N$ be the components of the state vector, and $c_N=\sqrt{\gamma p_N/\rho_N}$ the local speed of sound in the last computational cell. Assuming that the discharge proceeds isentropically and that the $I^+$ characteristic
\begin{equation}
I^+ = \frac{2}{\gamma-1}c+u
\end{equation}
transporting information out of the computational domain is conserved, we have for the values of the upper boundary (ub) cell:
\begin{eqnarray}
\rho_{ub} &=& \rho_N\left(\frac{p_{ub}}{p_N}\right)^{1/\gamma}\\
u_{ub} &=& \frac{2}{\gamma-1}\left(c_N - c_{ub}\right)+u_N\\
p_{ub} &=& 1.
\end{eqnarray}

At the bottom exit we impose the velocity $u_e$ with which the gas leaves the computational (Euler) domain. Since velocities between this point and the disc are small (see Section~\ref{sec:coupling}) conservation of mass allows us to calculate this velocity from the rate of change in cavity volume between the disc and the boundary cell, i.e. $u_e \cdot\pi \cdot r_e^2= \int_{S_b}\phi_n dS_b$ with the surface $S_b$ including the disc itself.
Similarly as at the upper exit, the state of the lower boundary (lb) cell can then be computed from the state of the first computational cell ($\rho_1$, $u_1$, $p_1$) using conservation of the $I^-$ characteristic:
\begin{eqnarray}
\rho_{lb} &=& \gamma \frac{p_{lb}}{c_{lb}^2}\\
u_{lb} &=& u_e\\
p_{lb} &=& \left[\left(\frac{\gamma}{c_{lb}^2}\right)^\gamma\frac{p_1}{\rho_1^\gamma}\right]^{1/(1-\gamma)}\\
c_{lb} &=& c_1 - \left(u_1 - u_e\right)\frac{\gamma-1}{2}.
\end{eqnarray}
These boundary conditions cannot completely prevent reflection of waves at the upper and lower ends which causes some small oscillations in the state variables $\rho$, $u$, and $p$ as can be seen for example in Fig.~\ref{fig:twoPhase}~(b). The oscillations however remain sufficiently small so that the averaged evolution of the gas dynamics can still reliably be extracted.

\subsection{Coupling between boundary-integral and Roe method}\label{sec:coupling}

\begin{figure}
\includegraphics[width=0.6\columnwidth]{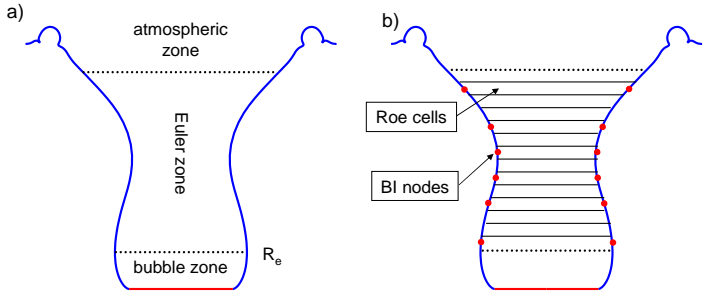}
\caption{(a) Illustration of the three zones in which the gas domain is split during the compressible stage: in the ``atmospheric zone'' on top, the pressure is always atmospheric and the gas dynamics are neglected. In the ``Euler zone'' we use the Roe scheme to calculate the fully compressible gas dynamics. In the ``bubble zone'' again the pressure is constant and given by the pressure at the end of the Euler zone. (b) Schematic illustration of the alignment of Euler cells and BI nodes: the BI nodes are always placed at the height of the center of the Euler cells. In the neck area a BI node is placed in every 2nd Roe cell, while away from the neck the spacing is larger to save computation time.}
\label{fig:illuZones}
\end{figure}

During the compressible stage of the simulation, the gas domain is split into three separate zones as illustrated in Fig.~\ref{fig:illuZones}~(a). The gas pressure which is required for coupling to the BIM is obtained in a different way for each zone. First, in the ``atmospheric zone'' the pressure is taken to be atmospheric since the gas dynamics is negligible. Next, in the ``Euler zone'' the pressure is provided by the solution of the Euler equations (\ref{eqn:mass})--(\ref{eqn:energy}). Finally, in the ``bubble zone'' the gas dynamics is again neglected. The pressure, however, cannot be taken to be atmospheric since the zone is not open to the atmosphere. Due to the low gas velocities in that zone (of order of the disc speed) the pressure is taken equal to the pressure at the bottom end of the Euler zone.

The upper end of the Euler zone can remain at a fixed vertical position (between 0 and 1 disc radii below the initial free surface). The bottom end is fixed to be at the maximum radial extension of the cavity below the neck as depicted in Fig.~\ref{fig:illuZones}~(a). Note that the Euler zone cannot be extended all the way down to the disc since the free surface departs almost horizontally from the disc's edge. This leads to large gradients in the cavity radius on the right-hand side of Eq.~(\ref{eqn:momentum}) and can thus cause numerical instabilities. Since the bottom end of the Euler zone is moving downward in time, its length needs to be extended during the simulation. Our algorithm adds a new Roe cell whenever the distance between the desired location (maximum radial extension of the cavity) and the actual position of the last Roe cell becomes larger than the size of a single cell.

For the results presented here, the initial number of cells in the Roe solver is 600 and grows dynamically by extension of the Euler zone. To ensure sufficient resolution for the high air speeds even during the last moments of the simulation, we double the number of cells by splitting each cell in half whenever the minimum cavity radius $r_\mathrm{neck}$ falls below a certain value. This is done twice: at $r_\mathrm{neck}=0.2$ and $r_\mathrm{neck}=0.05$. Due to extension and node doubling the number of Roe cells at the end of the simulation is somewhat above 3000. The height of the Roe cells is always constant with $\Delta z = 0.0012$. We find that the number of Roe cells is not crucial for the total computation time which is mainly determined by the BIM calculations.

It is crucial to properly align the positions of the BI nodes with the Euler cells. For this we place a BI node always exactly in the center of every $n$th Euler cell as illustrated in Fig.~\ref{fig:illuZones}~(b). Usually $n=2$ in a refined zone around the cavity neck and $n$=5 outside this zone. The cross-sectional area $S$ as well as its spatial and temporal derivatives for each cell are calculated at the height of the cell center from the splines interpolating the cavity surface in the BIM. To avoid numerical instabilities of the BIM we use the periodic regridding described above which now makes the BI nodes ``jump'' between Euler cells. This implies that there must always be at least one Roe cell without a BI node between two cells which contain a node, i.e.~$n\ge 2$.

The two-way coupling between the gas and the liquid domain is accomplished as follows.
At each time step $j$ we first do a BI step to advance the shape and velocity potential of the free surface from $j$ to $j+1$ using the gas pressure of step $j$. The BI step further provides the derivatives $\partial S/\partial z = 2 \pi r dr/dz$ and $\partial S / \partial t = 2\pi r \dot{r}$. This calculation is followed by a Roe step using the new cavity shape $j+1$ to obtain the new pressure at $j+1$ and so forth.

We performed an extensive set of simulations to verify that the results presented in the next section are numerically robust when changing any of the above mentioned simulation parameters.

%
%
%
%

\section{Results}

\subsection{Justification of the 1D compressible scheme}\label{sec:justification}

\begin{figure}
\begin{center}
\includegraphics[width=0.9\columnwidth]{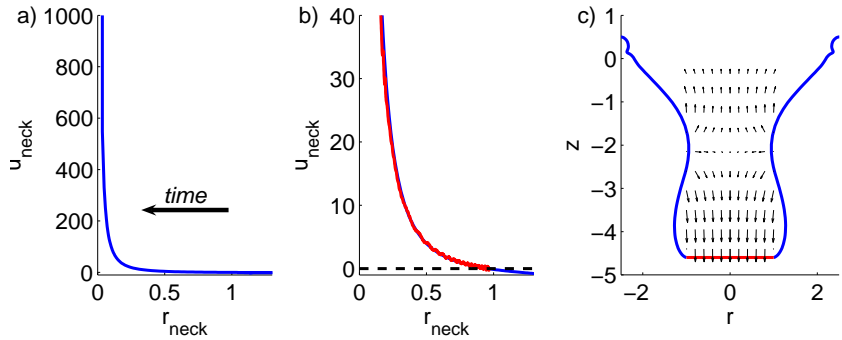}
\caption{(a) The gas velocity at the neck as obtained from an incompressible two-fluid BI simulation. The velocity easily surpasses the speed of sound demonstrating the need for our more sophisticated compressible approach. (b) The velocity obtained from the multiscale simulation (red line) agrees so well with the velocity from the two-fluid BIM (blue line) that both are hardly distinguishable for low velocities where compressibility is negligible. The start of the red curve marks the transition from the incompressible to the compressible stage. The small oscillations in the red curve are due to wave reflections at the ends of the Euler zone. (c) The gas flow field obtained from the two-fluid BIM at the moment when the velocity at the neck reverses and the simulation passes from the incompressible to compressible stage. Except for a rather narrow zone around the neck the flow is to a good approximation one-dimensional.}\label{fig:twoPhase}
\end{center}
\end{figure}

In Fig.~\ref{fig:twoPhase}~(a) we show the gas velocity through the cavity neck as a function of the shrinking cavity neck $r_\mathrm{neck}$ for the incompressible two-fluid BIM. We use $r_\mathrm{neck}$ instead of time to allow for an easier comparison with experiments in Section~\ref{sec:validation} and \cite{GekleEtAl_PRL_2010}. Already at $r_\mathrm{neck}=0.05$ (corresponding to 1~mm) an incompressible gas would surpass the speed of sound. This clearly demonstrates the need for a CFD method which takes the fully compressible gas dynamics into account if one wants to study the ejected air stream close to cavity collapse.

A full two-way coupling is required as both gas and liquid flows occur on similar time scales: We first estimate the typical time scale for the gas flow $T_\mathrm{gas}$ as the length of the cavity $L\approx 10$~cm divided by the speed with which a perturbation travels, $c=330$~m/s, to obtain $T_\mathrm{gas}=0.3$~ms. A typical time scale for the variation of the cavity radius can be derived by considering the time that it takes the cavity to collapse from a neck radius of 4~mm
\footnote{At $r_\mathrm{neck}\approx 4$~mm the vertical neck motion starts to reverse which marks the beginning when air effects become important, see Fig.~5 of \cite{GekleEtAl_PRL_2010}.}
down to zero which is $T_\mathrm{cav}\approx 1$~ms and thus of the same order as $T_\mathrm{gas}$.

For some time after switching from the incompressible to the compressible stage the gas velocities are still moderate and compressibility effects should be negligible. We can thus expect that in the beginning of the compressible stage the gas velocity obtained from our multiscale approach should be similar to the two-fluid BIM in Fig.~\ref{fig:twoPhase}~(a). That this is indeed the case is demonstrated in Fig.~\ref{fig:twoPhase}~(b) which gives us a first indication that the coupling between the Roe solver and the BIM works correctly. It further gives good evidence that the assumption of one-dimensional gas flow in the compressible stage is justified.

To verify the 1D assumption more explicitly, Fig.~\ref{fig:twoPhase}~(c) shows the flow field obtained from the two-fluid BIM at the moment of flow reversal. Except for a small region around the neck where by definition the vertical velocity is zero, our assumption is well justified.

\subsection{Structure of the compressible gas flow}\label{sec:structure}

\begin{figure}
\begin{center}
\includegraphics[width=0.5\columnwidth]{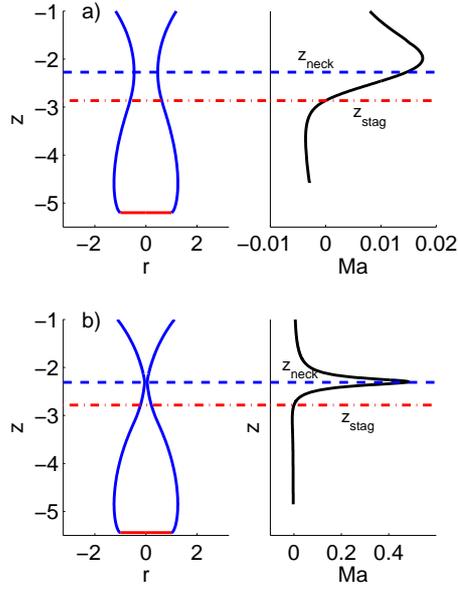}
\caption{The velocity of the gas stream in the Euler zone for various instants of the cavity collapse (right images) with corresponding cavity profiles (left images). (a) In the early stage for neck radii around half a disc radius ($r_\mathrm{neck}=0.46$) the velocity peak is still rather broad. At the bottom end gas leaves the Euler zone with a velocity approximately equal to the disc velocity of -1 which corresponds to Ma=-0.003. Note that the velocity peak is located somewhat upstream of the neck which is marked by the blue dashed line. This is markedly different from steady subsonic flow through a fixed nozzle where both would coincide. The location of the stagnation point is indicated by the red dash-dotted line. (b) At a later time ($r_\mathrm{neck}=0.05$) the velocity peak sharpens and increases in magnitude. The peak is now located almost at the neck.}\label{fig:velocityGas}
\end{center}
\end{figure}

The intricate structure of the gas flow in the Euler zone is illustrated by the velocity profile in Fig.~\ref{fig:velocityGas} for various instants of time. Here we normalize velocities with the speed of sound to obtain the Mach number Ma=$u/c$. At early times as in Fig.~\ref{fig:velocityGas}~(a) one appreciates that at the bottom end of the Euler zone air is entrained by the downward moving disc at velocities of the order of the disc speed. This downflux is however overcompensated by the shrinking of the cavity around the neck so that the total flux is directed upwards as can be seen by the velocity maximum at $z\approx -1.9$. Above the maximum the cavity widens again and the velocity decays towards the upper end of the cavity. The consequence of the competition between cavity expansion at the bottom and cavity collapse around the neck is the creation of a stagnation point with $\mathrm{Ma}=0$ as is clearly visible in Fig.~\ref{fig:velocityGas}~(a).

As the collapse progresses gas is pushed through the rapidly diminishing cavity neck at ever higher and higher speeds leading to a sharp velocity peak at the neck as illustrated in Fig.~\ref{fig:velocityGas}~(b). At the top and bottom boundaries of the Euler zone the velocities remain almost unaltered as compared to Fig.~\ref{fig:velocityGas}~(a).

\begin{figure}
\begin{center}
\includegraphics[width=0.3\columnwidth]{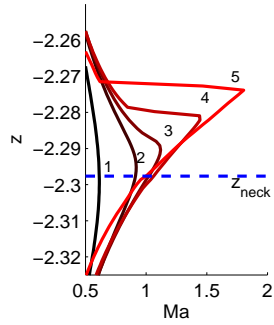}
\caption{A shock wave develops (the numbers 1--5 correspond to neck radii of 0.044, 0.036, 0.034, 0.032 and 0.027; the blue dashed line indicates the neck position for curve number 5). For even smaller neck radii the simulation destabilizes and no reliable data is available.}\label{fig:velocityGas_shock}
\end{center}
\end{figure}

Finally, as the flow speed increases even further a shock wave develops upstream of the neck as shown by the magnification in Fig.~\ref{fig:velocityGas_shock}. Thanks to the shock-capturing ability of the employed Roe scheme our method is able to handle the shock formation quite well.

\begin{figure}
\begin{center}
\includegraphics[width=0.5\columnwidth]{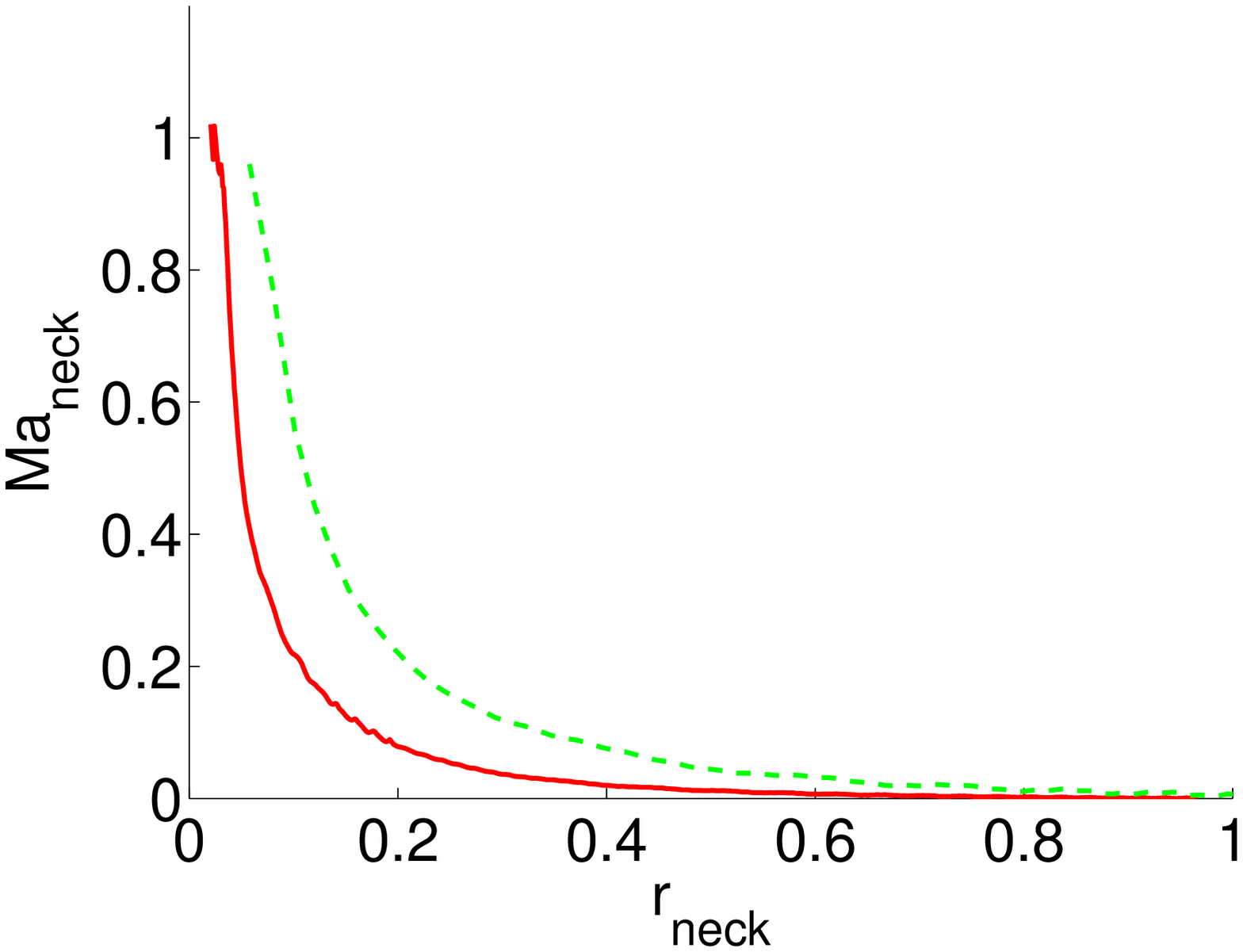}
\caption{
The Mach number at the neck for two different impact speeds: 1~m/s (red solid curve) and 2~m/s (green dashed curve). The flow becomes sonic when neck has shrunk to 0.025 or 0.06 disc radii, respectively.}\label{fig:machNeck}
\end{center}
\end{figure}

Figure~\ref{fig:machNeck} shows the Mach number at the cavity neck. For our standard configuration of a 2~cm disc impacting at 1~m/s the flow becomes sonic at a neck radius of 0.025 (corresponding to 0.5~mm). Here we also add data for a higher impact speed of 2~m/s which qualitatively shows the same behavior but where sonic flow is attained already at a neck radius of 0.06 (1.2~mm). Once sonic velocities at the neck are reached our numerical scheme becomes unstable which is why do not present any data beyond this point. Note that due to our unsteady situation supersonic speeds are attained even earlier at locations above the neck, cf.~Fig.~\ref{fig:velocityGas_shock}.

\begin{figure}
\begin{center}
\includegraphics[width=0.5\columnwidth]{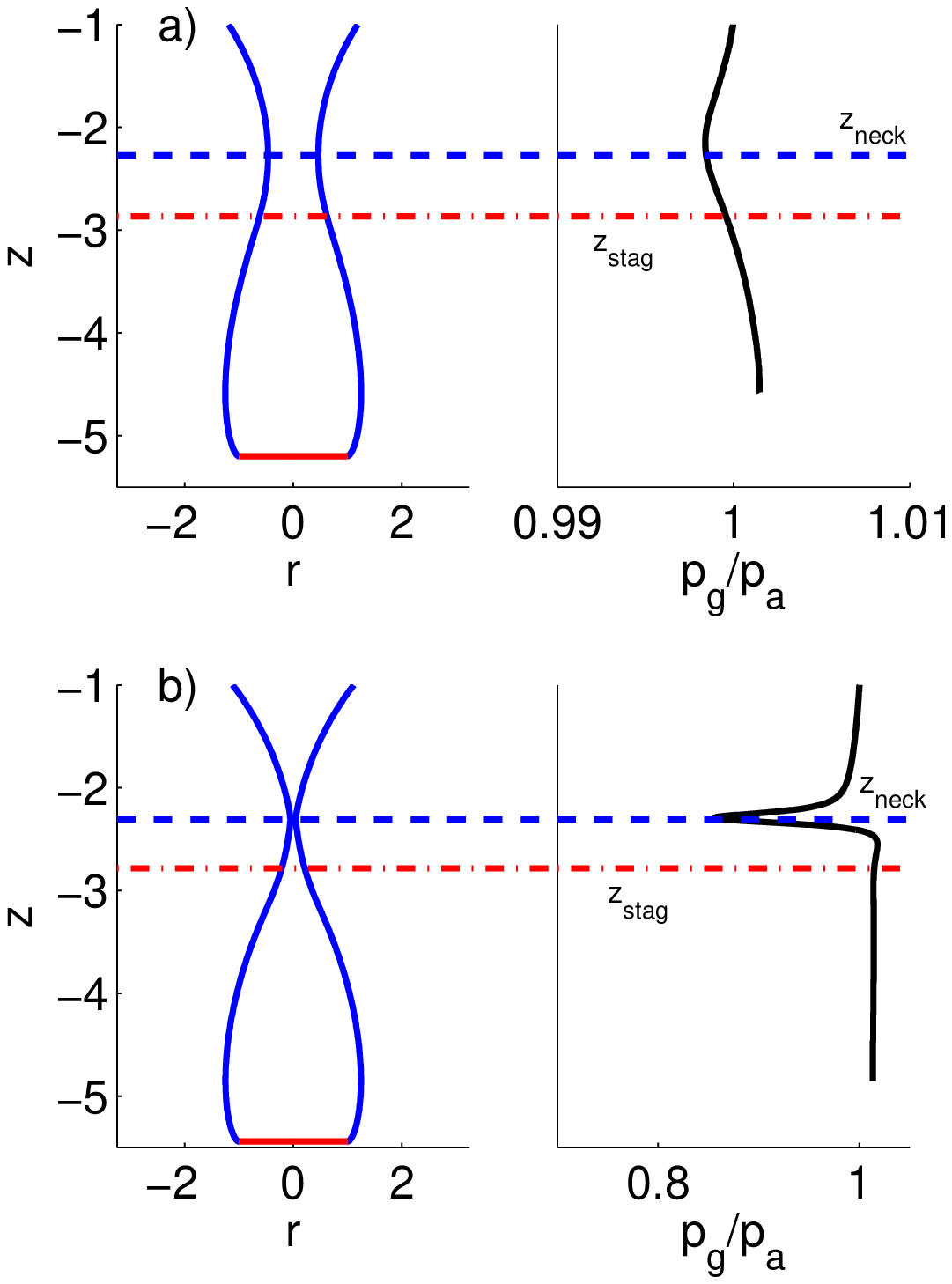}
\caption{The pressure profile in the Euler zone at the same instants as in Fig.~\ref{fig:velocityGas}. The low pressure at the neck is caused by the high gas speeds in that region.}\label{fig:pressureGas}
\end{center}
\end{figure}

\begin{figure}
\begin{center}
\includegraphics[width=0.7\columnwidth]{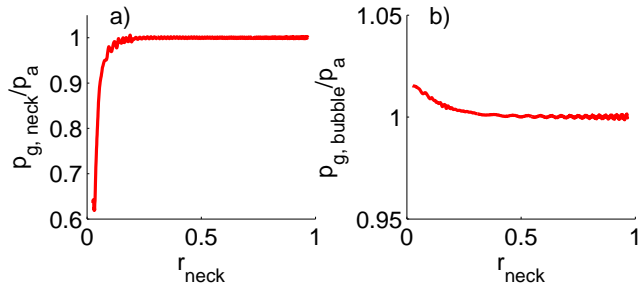}
\caption{(a) The pressure at the neck diminishes over time to reach a minimum value of approx.~0.6 atmospheres. (b) In contrast, the pressure deep inside the bubble rises only slightly above atmospheric towards the end.}\label{fig:pressureGas_bubbleNeck}
\end{center}
\end{figure}

We now turn to study the pressure distribution in the Euler zone which is illustrated in Fig.~\ref{fig:pressureGas}. In the early stages of the process (Fig.~\ref{fig:pressureGas}~(a)) the pressure remains virtually atmospheric with only a very slight dip around the cavity neck. At a later time, however, the pressure at the neck diminishes substantially as can be seen in Fig.~\ref{fig:pressureGas}~(b). In a steady state situation one would expect the neck pressure to reach a minimum value of
\begin{equation}
p_\mathrm{neck} = \left(1+\frac{\gamma-1}{2}\right)^{-\gamma/(\gamma-1)} \approx 0.53
\end{equation}
with $\gamma=1.4$ the isentropic exponent, as $\mathrm{Ma_{neck}}$ becomes unity. As shown in Fig.~\ref{fig:pressureGas_bubbleNeck}~(a) our situation -- although highly unsteady -- exhibits a similar behavior with $p_\mathrm{neck}\approx 0.6$ at the final instant before the simulation destabilizes.

Notably, below the neck the pressure is very uniform all the way down to the end of the Euler zone (Fig.~\ref{fig:pressureGas_bubbleNeck}~(a)) which allows us to define a single pressure value for the ``bubble'' between the neck and the disc. Figure~\ref{fig:pressureGas_bubbleNeck}~(b) demonstrates that in our unique situation of air being pushed through a rapidly shrinking nozzle sonic speeds can be attained with a bubble pressure which is merely 2\% higher than the surrounding atmosphere.

\subsection{Experimental validation}\label{sec:validation}

Here we describe briefly the validation of our numerical results with the experimental data of \cite{GekleEtAl_PRL_2010} which is achieved in three different ways:

(i) We use smoke particles to directly measure the air speed as it is pushed out of the collapsing cavity. We can reliably measure air speeds up to 10~m/s and very good agreement with the numerical data is found as presented in Fig.~2 of \cite{GekleEtAl_PRL_2010}.

(ii) In order to confirm the validity of our predictions also for higher gas velocities we compare the numerical cavity shape close to pinch-off with experimental images. We find that the experimental shape is not smoothly curved but exhibits a rather pronounced ``kink'' at its neck. This effect -- which is due to the low pressure induced by the high gas speeds -- is not present in single-fluid simulations, but can be reproduced rather well by the inclusion of air effects as shown in Fig.~3 of \cite{GekleEtAl_PRL_2010}.

(iii) Finally, we show that the cavity neck in the experiment exhibits a significant upward motion prior to final collapse as the fast air stream pushes the surface minimum upwards. We find very good agreement for this motion between experiment and our multiscale simulations as shown in Fig.~4 of \cite{GekleEtAl_PRL_2010}.

The quantitatively consistent observation of the above air effects in our compressible simulations and corresponding experiments gives us strong confidence in the reliability of our numerical scheme.

\section{Conclusions}\label{sec:conclusions}

We presented a multiscale model to simulate the impact of a solid object onto a liquid surface. Our focus was on the fast stream of air that is pushed upwards as the impact cavity collapses due to hydrostatic pressure. We showed that in our case of a 2~cm disc impacting at 1~m/s the air attains supersonic velocities and thus requires the use of a fully compressible computational method -- in contrast to existing treatments such as Volume-of-Fluid or Level-Set methods \cite{JosserandZaleski_PhysFluids_2003, LeneweitEtAl_JFM_2005, ZhengEtAl_JComputPhys_2005, DoQuangAmberg_PhysFluids_2009, CoyajeeBoersma_JComputPhys_2009} in which the air flow was considered incompressible.

In our simulations the impact process is split in an incompressible and a compressible stage. During the incompressible stage which covers the first part of the process, air is entrained into the cavity at relatively low speeds (compared to the speed of sound). This allows us to use a two-fluid boundary-integral method for the gas and the liquid domain. The compressible stage starts as the air flow reverses and air is pushed out through the cavity neck. In this stage we couple two different methods: a Roe scheme to solve the one-dimensional Euler equations in the gas domain and a single-fluid boundary-integral method in the liquid domain. The two domains are connected via the pressure at the free surface.

The predominantly one-dimensional character of the air stream and the shape of the impact cavity make our system reminiscent to the common problem of air flow through a converging-diverging nozzle in aerodynamics. There is, however, an important and fundamental difference: since in our case the confining cavity is a \emph{liquid}, the ``nozzle'' shape is rapidly evolving in time.

For low gas velocities we find very good agreement between our simulations and direct experimental measurements \cite{GekleEtAl_PRL_2010}. The shape of the cavity as well as a final upward motion of the cavity neck \cite{GekleEtAl_PRL_2010} give further strong evidence that our multiscale numerical method faithfully reflects reality.

\begin{acknowledgements}
We thank Ivo Peters for providing experimental data and Devaraj van der Meer as well as Detlef Lohse for discussions. This work is part of the program of the Stichting FOM, which is financially supported by NWO. JMG thanks the financial support of the Spanish Ministry of Education under project DPI2008-06624-C03-01.
\end{acknowledgements}


\begin{thebibliography}{10}
\providecommand{\url}[1]{{#1}}
\providecommand{\urlprefix}{URL }
\expandafter\ifx\csname urlstyle\endcsname\relax
  \providecommand{\doi}[1]{DOI \discretionary{}{}{}#1}\else
  \providecommand{\doi}{DOI \discretionary{}{}{}\begingroup
  \urlstyle{rm}\Url}\fi

\bibitem{Worthington_book_1908}
A.M. Worthington, \emph{A study of splashes} (Longmans, Green and Co., London,
  1908)

\bibitem{OguzProsperetti_JFM_1990}
H.N. Oguz, A.~Prosperetti, J.~Fluid~Mech. \textbf{219}, 143 (1990)

\bibitem{MortonRudmanLiow_PhysFluids_2000}
D.~Morton, M.~Rudman, J.L. Liow, Phys.~Fluids \textbf{12}, 747 (2000)

\bibitem{JosserandZaleski_PhysFluids_2003}
C.~Josserand, S.~Zaleski, Phys.~Fluids \textbf{15}, 1650 (2003)

\bibitem{LeneweitEtAl_JFM_2005}
G.~Leneweit, R.~Koehler, K.G. Roesner, G.~Sch{\"a}fer, J.~Fluid~Mech.
  \textbf{543}, 303 (2005)

\bibitem{ZhengEtAl_JComputPhys_2005}
X.~Zheng, J.L.A. Anderson, V.~Cristini, J.~Comput.~Phys. \textbf{208}, 626
  (2005)

\bibitem{CoyajeeBoersma_JComputPhys_2009}
E.~Coyajee, B.J. Boersma, J.~Comput.~Phys. \textbf{228}, 4444 (2009)

\bibitem{GreenhowMoyo_PhilTransRoySocLondonA_1997}
M.~Greenhow, S.~Moyo, Phil.~Trans.~R.~Soc.~London~A \textbf{355}, 551 (1997)

\bibitem{Gaudet_PhysFluids_1998}
S.~Gaudet, Phys.~Fluids \textbf{10}, 2489 (1998)

\bibitem{BattistinIafrati_JFluidStruct_2003}
D.~Battistin, A.~Iafrati, J.~Fluid~Struct. \textbf{17}, 643 (2003)

\bibitem{LiEtAl_JComputPhys_2005}
J.~Li, M.~Hesse, J.~Ziegler, A.W. Woods, J.~Comput.~Phys. \textbf{208}, 289
  (2005)

\bibitem{VellaMetcalfe_PhysFluids_2007}
D.~Vella, P.D. Metcalfe, Phys.~Fluids \textbf{19}, 072108 (2007)

\bibitem{Lin_ComputFluids_2007}
P.~Lin, Comput.~Fluids \textbf{36}, 549 (2007)

\bibitem{GekleEtAl_PRL_2008}
S.~Gekle, A.~van~der Bos, R.~Bergmann, D.~van~der Meer, D.~Lohse,
  Phys.~Rev.~Lett. \textbf{100}, 084502 (2008)

\bibitem{GekleEtAl_PRL_2009}
S.~Gekle, J.M. Gordillo, D.~van~der Meer, D.~Lohse, Phys.~Rev.~Lett.
  \textbf{102}, 034502 (2009)

\bibitem{BergmannEtAl_JFM_2009}
R.~Bergmann, D.~van~der Meer, S.~Gekle, A.~van~der Bos, D.~Lohse,
  J.~Fluid~Mech. \textbf{633}, 381 (2009)

\bibitem{DoQuangAmberg_PhysFluids_2009}
M.~Do-Quang, G.~Amberg, Phys.~Fluids \textbf{21}, 022102 (2009)

\bibitem{GekleEtAl_PRL_2010}
S.~Gekle, I.R. Peters, J.M. Gordillo, D.~van~der Meer, D.~Lohse,
  Phys.~Rev.~Lett. \textbf{104}, 024501 (2010)

\bibitem{BartoloJosserandBonn_PRL_2006}
D.~Bartolo, C.~Josserand, D.~Bonn, Phys.~Rev.~Lett. \textbf{96}, 124501 (2006)

\bibitem{DengAnilkumarWang_JFM_2007}
Q.~Deng, A.V. Anilkumar, T.G. Wang, J.~Fluid~Mech. \textbf{578}, 119 (2007)

\bibitem{GekleGordillo_preprint}
S.~Gekle, J.M. Gordillo, arXiv:0907.5154v1 [physics.flu-dyn]  (2009)

\bibitem{ChengLiu_JHydrodynamicsB_2007}
Y.S. Cheng, H.~Liu, J.~Hydrodynamics~Ser.~B \textbf{19}, 403 (2007)

\bibitem{Laney_book_1998}
C.~Laney, \emph{Computational gasdynamics} (Cambridge University Press, 1998)

\bibitem{Roe_JComputPhys_1981}
P.L. Roe, J.~Comput.~Phys. \textbf{43}, 357 (1981)

\bibitem{GodderidgeEtAl_OceanEng_2009}
B.~Godderidge, S.~Turnock, C.~Earl, M.~Tan, Ocean Eng. \textbf{36}, 578 (2009)

\bibitem{ChenPrice_PhysFluids_2009}
Y.G. Chen, W.G. Price, Phys.~Fluids \textbf{21}, 112105 (2009)

\bibitem{Pozrikidis_book_1997}
C.~Pozrikidis, \emph{Introduction to theoretical and computational fluid
  dynamics} (Oxford University Press, 1997)

\bibitem{BlakeTaibDoherty_JFM_1986}
J.R. Blake, B.B. Taib, G.~Doherty, J.~Fluid~Mech. \textbf{170}, 479 (1986)

\bibitem{OguzProsperetti_JFM_1993}
H.N. Oguz, A.~Prosperetti, J.~Fluid~Mech. \textbf{257}, 111 (1993)

\bibitem{RodriguezRodriguezGordilloMartinezBazan_JFM_2006}
J.~Rodr\'iguez-Rodr\'iguez, J.M. Gordillo, C.~Mart\'inez-Baz\'an,
  J.~Fluid~Mech. \textbf{548}, 69 (2006)

\bibitem{GordilloSevillaMartinezBazan_PhysFluids_2007}
J.M. Gordillo, A.~Sevilla, C.~Mart\'inez-Baz\'an, Phys.~Fluids \textbf{19},
  077102 (2007)

\end{thebibliography}

\end{document}